\RequirePackage{lineno}
\documentclass[aps,prl,preprint,superscriptaddress]{revtex4}

\usepackage{graphicx}
\usepackage{bm}
\usepackage{amsmath}
\usepackage{amssymb}
\usepackage{hyperref}

\begin{document}

\title{Optimizing phonon scattering by tuning surface-interdiffusion-driven intermixing to break the random-alloy limit of thermal conductivity}

\author{Xiaolong Yang}
\affiliation{Frontier Institute of Science and Technology, and State Key Laboratory for Mechanical Behavior of Materials, Xi'an Jiaotong University, Xi'an 710049, P. R. China.}

\author{Wu Li}
\email{wu.li.phys2011@gmail.com}
\affiliation{Institute for Advanced Study, Shenzhen University, Nanhai Avenue 3688, Shenzhen 518060, P. R. China.}

\begin{abstract}

We investigate the evolution of the cross-plane thermal conductivity $\kappa$ of superlattices (SLs) as interfaces change from perfectly abrupt to totally intermixed, by using non-equilibrium molecular dynamics (NEMD) simulations in combination with the spectral heat current calculations (SHCC). We highlight the role of surface-interdiffusion-driven intermixing by calculating the $\kappa$ of SLs with changing interface roughness, whose tuning allows for the $\kappa$ values much lower than the ``alloy limit'' and the abrupt interface limit in same cases. The interplay between alloy and interface scattering in different frequency ranges provides a physical basis to predict a minimum of thermal conductivity. More specifically, we also explore how the interface roughness affects the thermal conductivities for SLs materials with a broad span of atomic mass and bond strength. In particular, we find that (i) only when the ``spacer'' thickness of SLs increases up to a critical value the $\kappa$ of rough SLs can break the corresponding ``alloy limit'', since SLs with different ``spacer'' thickness have different characteristic length of phonon transport which are influenced by surface-interdiffusion-driven intermixing to different extend. (ii) Whether the $\kappa$ changes monotonically as interface roughness strongly depends on the period length and intrinsic behavior of phonon transport for SLs materials. Especially, for the SL with large period length, there exists an optimal interface roughness which can minimize the thermal conductivity. (iii) Surface-interdiffusion-driven intermixing is more effective in achieving the low $\kappa$ below the alloy limit for SL materials with large mass mismatch than with small one. (iv) It's possible for SLs materials with large lattice mismatch (i.e., bond strength) to design an ideally abrupt interface structure with $\kappa$ much below the ``alloy limit''. These results have a clear implications for optimization of thermal transport for heat management and for the development of thermoelectric materials.
\end{abstract}

\maketitle
%\linenumbers

\section{Introduction}

Accurately manipulating the thermal conductivity is a fundamental challenge for many technologies including phase-change memory development, micro- and nanoelectronics heat management, and thermoelectricity \cite{Chowdhury2009, Cahill2014}. In particular, thermoelectric materials capable of converting heat into electric power and vice versa, have attracted increasing interest for applications in energy harvesting and interconnection technologies\cite{Nielsch2011,Zebarjadi2012}. Its efficiency has, however, been hindered on finding materials with low thermal conductivities $\kappa$. Given that nanostructuring enables dramatic reductions of thermal conductivities by scattering phonons at nanscale interface and defects, and thus is considered as an effective strategy to enhance the thermoelectric efficiency \cite{Liu2011, Zebarjadi2011, Chen2012}. A typical approach involves interfaces in superlattices (SLs), which have emerged so far as a promising strategy to low the thermal conductivity by interface scattering. To this end, many efforts have been dedicated to the investigation of the effects of roughness \cite{Tian2012,Stoltz2009}, interdiffusion \cite{Chen2013}, lattice mismatch \cite{Obrien2013,Wang2016,Mizuno2015,Guo2015Approaching}, coherence \cite{Luckyanova2012}, and nanoscale constrictions \cite{Prasher2005} on the cross-plane thermal conductivity of superlattices. Among these, understanding the effect of interface roughness driven by interdiffusion on thermal transport is increasingly crucial for achieving highly diffusive phonon scattering at the interfaces, and thus become a focus of attention in the search of efficient nanostructured thermoelectric materials.

Only very recently has interdiffusion around the interfaces been reported in a few of the experimental studies \cite{Koester2001,Magri2002,Aubertine2005,Stoltz2009,Chen2013,Chen2015}. For instance, Chen and co-workers demonstrated by using a combination of experiment and atomistic ab initio calculations that Ge-segregation-driven intermixing around the interfaces is able to lower $\kappa$ below both the alloy limit and the abrupt interface limit in SiGe superlattices\cite{Chen2013}. Recently, they also investigated the evolution of structure and cross-plane thermal conductivity $\kappa$ of Ge/Si SLs induced by post-growth annealing using transmission electron microscopy (TEM), and they demonstrated that phonon scattering by the interfaces can be suppressed and eliminated by enhancing Si-Ge intermixing around the interfaces \cite{Chen2015}. Although these previous results revealed that Ge-segregation-driven intermixing can lead to $\kappa$ values much lower than the ``alloy limit'', precise control of the interface roughness by experimentally thermal annealing is so far limiting, and thus it is in fact still not very clear how the $\kappa$ of a SL changes with interface roughness driven by surface interdiffusion.

With the motivation above, we perform non-equilibrium molecular dynamics (NEMD) simulations to investigate the evolution of the cross-plane thermal conductivity  of superlattices as interfaces change from perfectly abrupt to totally intermixed. To address this issue, we investigate the thermal transport in the binary Lennard-Jones SLs with different interface roughness, and the corresponding alloy systems consisting of two base materials with different atomic mass or bond strength. We demonstrate that surface-interdiffusion-driven intermixing is very crucial for achieving the low $\kappa$, especially, when the period thickness of the base material with a light atomic mass increases up to a critical value the $\kappa$ of rough SLs can break the corresponding ``alloy limit''. In addition, by calculating the $\kappa$ of SLs with increasing interface roughness, we find the $\kappa$ does not monotonically changes with interface roughness for the SLs with large period length. Instead, there exists an intermediate interface roughness which can minimize the thermal conductivity. From applied point of view, we also study the effect of interface roughness on the thermal conductivities for SLs materials with a broad span of atomic mass and bond strength. Our simulation results shows that surface-interdiffusion-driven intermixing is more effective in achieving the low $\kappa$ below the alloy limit for SL materials with large mass mismatch than with small one. Furthermore, we find it possible for perfectly SLs materials with large lattice mismatch (i.e., bond strength) to design an ideally abrupt interface structure with $\kappa$ much below the ``alloy limit''. The remainder of this paper is organized as follows. In Sec. II we describe the model system and the setup of our NEMD simulations. Then we present the NEMD simulation results and our discussions in Sec. III. Finally, we conclude this paper in Sec.IV.

\section{Methodology}
\subsection{ Model system}

The model structures of SLs and random alloy are constructed via layer-by-layer stacking of face-centered-cubic (FCC) unit cells (UC) of two different materials A and B alternatively along the [100] direction. To clarify, for the sake of generality, herein A is a type of silicon-like material with light atomic mass or strong bond strength called ``spacer'' in which phonons travel ballistically, and accordingly, B represents the germanium-like material with heavy atomic mass or weak bond strength called ``barrier'' where phonons are scattered in a completely diffusive way, as defined by Ref.\cite{Chen2013}. Specifically, the SL consists of N periods of n monolayers (ML) of A separated by m ML of B, as shown in Fig.\ref{fig:1}(a). Herein, all the interatomic interactions are described by the Lennard-Jones (LJ) potential,

\begin{equation}
\label{eq:1}
\phi_{ij}(r_{ij})=4\epsilon \left[\left(\frac{\sigma}{r_{ij}}\right)^{12} -\left(\frac{\sigma}{r_{ij}}\right)^6\right],
\end{equation}

\noindent where $\phi_{ij}$ and $r_{ij}$ are the pairwise interaction potential energy and the distance between atoms i and j, $\sigma$ is the pair separation at which potential energy is zero, and $\epsilon$ is the depth of potential well, respectively. While LJ potential cannot provide an adequate quantitative description of real materials, it is undeniable that LJ systems have been widely applied to mimic materials with a wide span of atomic mass and bond strength \cite{Stevens2007,Lyver2009,Landry2009,Roberts2010} due to both computationally inexpensive cost and the simplicity in controlling the lattice spacing (by $\sigma$) and the interatomic interaction strength (by $\epsilon$), i.e., Si/Ge\cite{Borca-Tasciuc2000,McGaughey2006}, AlAs/GaAs\cite{Capinski1999}, Bi$_2$Te$_3$/Sb$_2$Te$_3$\cite{Capinski1999}. The parameters set for all the atomic interactions in this study is $\sigma = 0.34nm$, $\epsilon = 0.1664eV$, and a cutoff radius of 2.5$\sigma$ same as that in solid argon provided in previous studies \cite{Wang2014,Lyver2009,Wang2015}. It should be noted that the only difference between the two base materials A and B is their atomic masses, which are, respectively, $m_{A}$=40g/mol and $m_{B}$= 90g/mol corresponding to the case of the realistic Si-Ge superlattice unless otherwise mentioned.

\subsection{Interfacial species mixing}

\begin{figure}[htp]
	\centerline{\includegraphics[width=13cm]{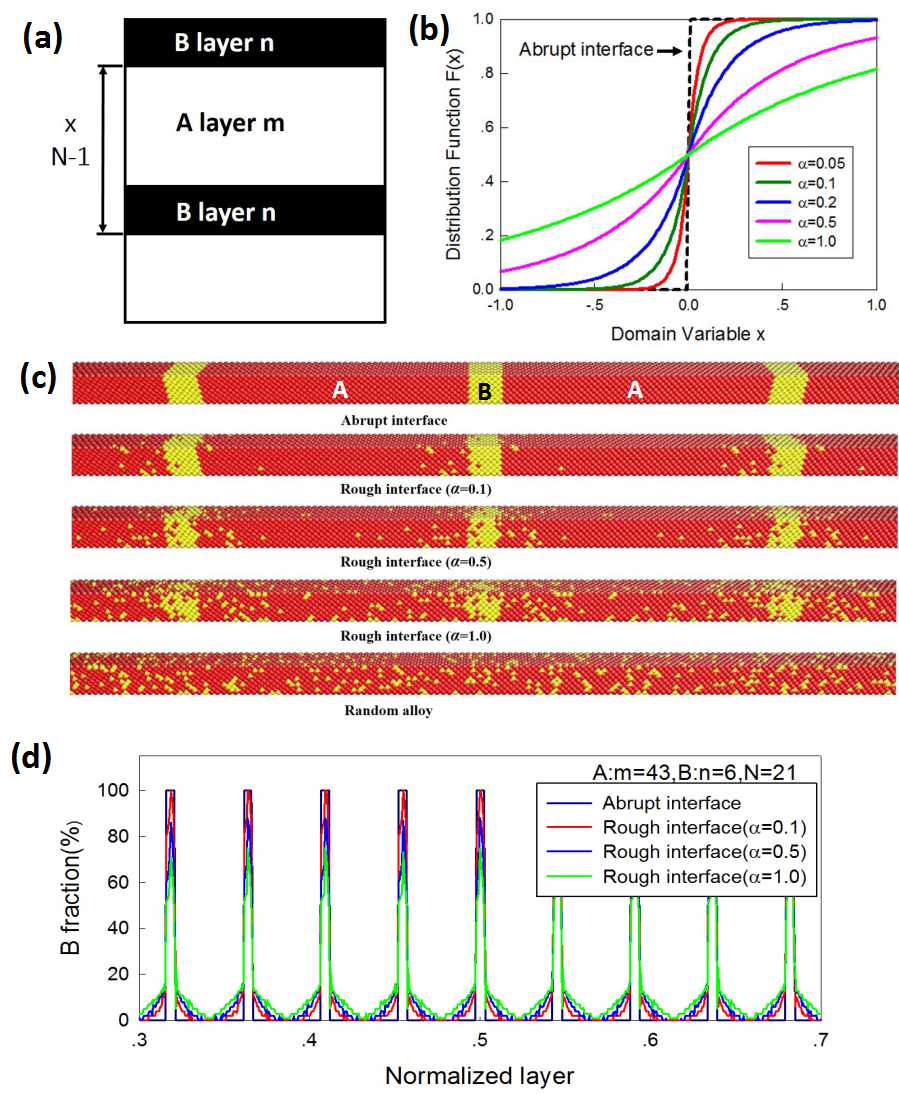}}
	\caption{\label{fig:1}
	(a) Schematic illustration of $(A)_m/(B)_n$ SL structures used for our simulation. N denotes the period numbers of n monlayers (ML) of B separated by m ML of A. (b) The log-ratio Laplace distribution function for various values of $\alpha$. (c) Snapshots of the atomistic structures as interfaces change from atomically sharp to totally intermixed. From up to down: perfectly abrupt interface, rough interface, random alloy. (d) Calculated B-concentration profiles corresponding to the SL structures given by (c). }
\end{figure}

In order to mimic the experimentally observed superlattices with surface-interdiffusion-driven intermixing\cite{Koester2001,Magri2002,Aubertine2005}, here a standard Laplace distribution function $F(x)$ was introduced to statistically describe the probability that an atomic site is occupied by B atom type over the entire SL structure\cite{English2012,Landry2009}

\begin{equation}
 \label{eq:2}
  F(x)=
  \begin{cases}
  \frac{1}{2}\exp(\frac{x-\beta_i}{\alpha}) &\mathrm{if} \quad x <= \beta_i,\\
  1- \frac{1}{2}\exp(-\frac{x-\beta_i}{\alpha}) &\mathrm{if} \quad  x > \beta_i,
  \end{cases}
\end{equation}

\noindent where $x$ is measured relative to the closest interface, $\beta_i$ is a coordinate of the ith interface location, and $\alpha$ is a scale parameter used to describe the degree of the atomic intermixing around the interface ($\alpha \textgreater 0$). Atomic species mixing in cross section  are included by randomly assigning the species of each atom. Plots of the log-ratio Laplace distribution function for several values of $\alpha$ are shown in Fig.\ref{fig:1}(b). We can easily construct the atomistic structures of SL with different interface roughness according to this distribution by adjusting the values of $\alpha$. Fig.\ref{fig:1}(c) shows the examples of the constructed SL structures with different interface roughness, and the corresponding B-concentration profiles are also shown in Fig.\ref{fig:1}(d).

\subsection{Non-equilibrium molecular dynamics}

In our simulations, thermal conductivities are calculated using non-equilibrium molecular dynamics (NEMD) simulations \cite{Landry2009,Yang2015a,Xiong2016} as illustrated in the Fig.\ref{fig:2}(a), in which a temperature difference is imposed between the two ends of the simulation domain. The temperature gradient $\partial T/\partial x$, as well as the heat flux $J$, are measured after the system reached the stationary state. The temperature profiles of a SL with perfect interfaces and a SL with rough interfaces and the corresponding alloy are shown in Fig.\ref{fig:2}(b) as examples. The thermal conductivity $\kappa$ is then extracted from Fourier's law

\begin{equation}
\label{eq:3}
\kappa=-\frac{J}{\partial T/\partial x}.
\end{equation}

To carry out the NEMD simulations, we employed the classical parallel molecular dynamics package LAMMPS \cite{Plimpton1993}with a velocity Verlet algorithm for numerical integration of the equations of motion and a time step $dt$=1fs \cite{Yang2015a,Yang2015,Zhu2015}. We applied periodic boundary conditions in the transverse ($y$ and $z$) directions and fixed boundary conditions in the transport ($x$) direction. More specifically, the system is first relaxed in the isothermal-isobaric (NPT) ensemble using the Nose-Hoover thermostat for $5\times 10^6$ time steps. We then switch to the canonical (NVE) ensemble for a further $10^8$ time steps to gather the required statistics. The temperatures at the source and sink regions of the simulation domain were controlled with Langevin thermostats. In the present work, the simulation is primarily made for T=30K, and the temperature dependence of the thermal conductivity is also investigated.

\subsection{Spectral heat current calculation from NEMD simulations}

The phonon transmission functions across the SL interfaces are calculated based on the spectral heat current method developed by Saaskilahti \cite{Saeaeskilahti2015,PhysRevE.93.052141, PhysRevB.90.134312}. Considering the harmonic effect only, the frequency-dependent spectral heat current across the interface is given by the expression

\begin{equation}
\label{eq:4}
 q_{i\rightarrow j} = -\frac{2}{t \omega} \sum_{a,b \in {x,y,z}} \mathrm{Im} \langle v_{i}^{a}(b)^* K_{ij}^{a b} v_{j}^{b}(\omega)  \rangle,
\end{equation}

 \noindent where $\omega$ and $t$ are the angular frequency and the simulation time, respectively. $v_{i}^{a} (\omega)$  and $v_{j}^{b} (\omega)$ are the Fourier transformation of atomic velocities of atom i in $a$ direction and atom j in $b$ direction, respectively. $K_{ij}$ represents the harmonic spring constant between atom i in $a$ direction and atom j in $b$ direction, which corresponds to the second order derivatives of the interatomic potential energy with respect to the displacements around the equilibrium positions. Full expressions for all terms are not included here for brevity; readers are referred to Ref.\cite{Saeaeskilahti2015, PhysRevB.90.134312}for details. The total heat current across the interface separating adjacent atom sets left($L$) and right ($R$) can be obtained by summing over atoms in each set:

\begin{equation}
\label{eq:5}
 q(\omega)=\sum_{i\in L} \sum_{j\in R} q_{i\rightarrow j} (\omega).
\end{equation}

\begin{figure}[htp]
	\centerline{\includegraphics[width=15cm]{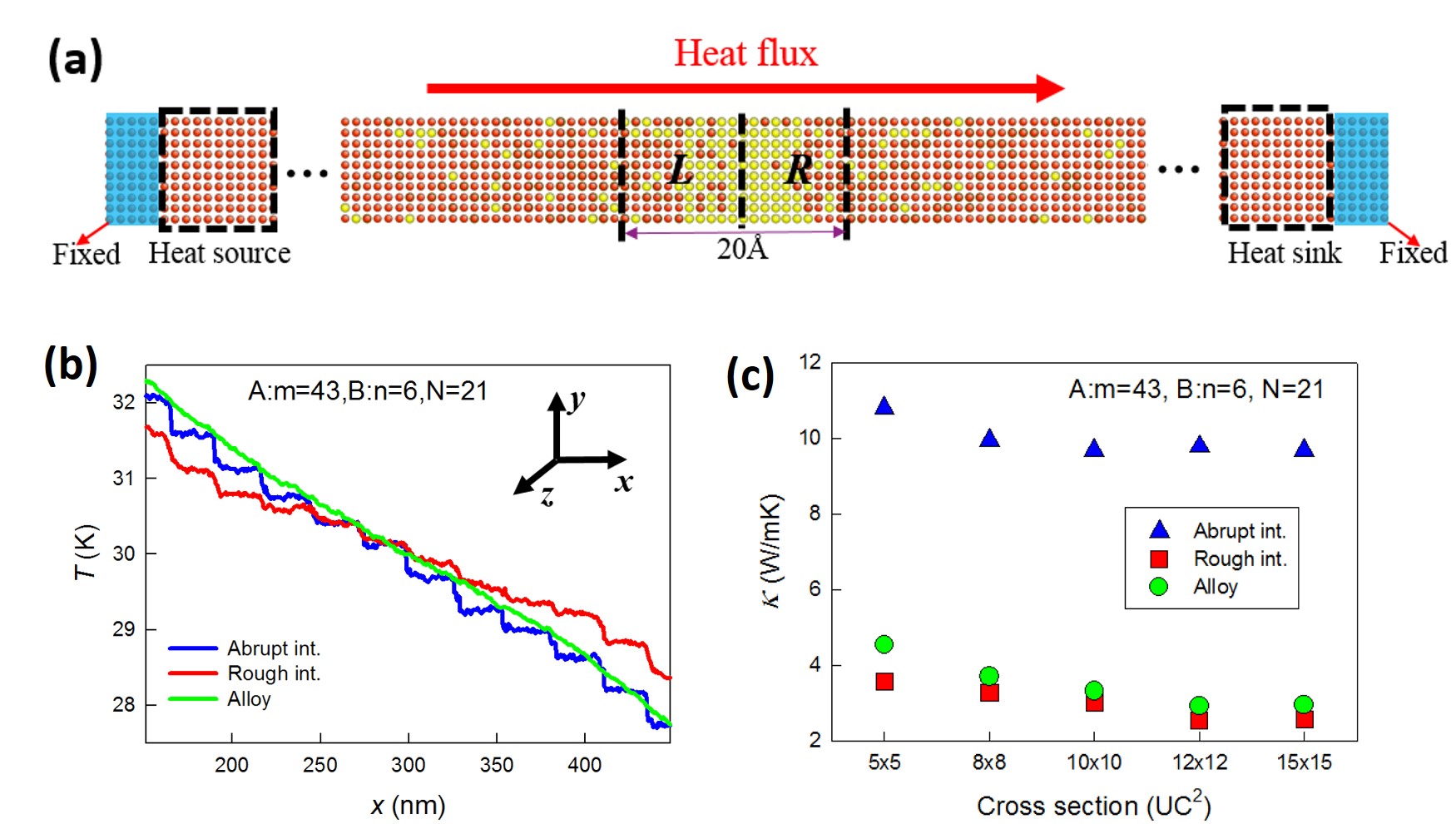}}
	\caption{\label{fig:2}
		(a) Schematic illustration of the phonon transmission function calculation set up from NEMD simulations. (b) Temperature profiles of the sample region along the transport ($x$) direction. (c) Dependence of the cross-sectional area on the conductivity for a SL with sharp interface, SL with interface roughness, and corresponding alloy, respectively. }
\end{figure}

Knowledge of $q(\omega)$ is enough to evaluate the accumulation of spectral heat current $Q$, defined as the integrated spectral heat current $q(\omega)$ up to $\omega_{max}$, which can be expressed as:

\begin{equation}
\label{eq:6}
Q=\int_0^{\omega_{max}}q(\omega)d\omega,
\end{equation}

\noindent where $\omega_{max}$ is the upper limit of phonon angular frequency. It is important to point out that the contribution of anharmonic effects is confirmed to be negligible by comparing the $Q$ to the total heat flux $J$ determined from the NEMD by the heat baths where full anharmonic effects are intrinsically included. As can be clearly seen from Fig.\ref{fig:5}(b), the difference within $6\%$ between the $Q$ and $J$ ensures that harmonic effects dominates the entire thermal transport in our systems, and thus we just consider the contribution of elastic interactions to the spectral heat current here for simplicity. After the knowledge of the above spectral heat current $q(\omega)$, the phonon transmission function across the interface can be thereby defined as

\begin{equation}
\label{eq:7}
\Gamma (\omega) = \frac{q(\omega)}{k_{B} \Delta T},
\end{equation}

\noindent where $\Delta T$ is the temperature difference between the two thermal baths in NEMD simulations, and $K_{B}$ is the Boltzmann constant. From the spectral heat current $q(\omega)$, we can also calculate the spectral decomposition of thermal conductivity across an individual interface as

\begin{equation}
\label{eq:8}
\kappa_{\mathrm{sp}} (\omega) = \frac{q(\omega)}{A \nabla T},
\end{equation}

 \noindent here A is the cross-sectional area of the simulation domain, and $\nabla T$ is the temperature gradient near the interface. The detailed simulation setup is schematically illustrated in Fig.\ref{fig:2}(a). The atoms at the two ends of SLs are fixed. Adjacent to the fixed layers, the atoms within the length $L_{bath}$ in the left side and right side are coupled to source and sink Langevin heat baths at temperatures $T+\Delta T/2$ and $T-\Delta T/2$, respectively. $2$ns long simulations were performed to gather the atomic velocities after the system reached the non-equilibrium steady state. Fig.\ref{fig:2}(c) shows the effect of cross-sectional area A on the thermal conductivity prediction for a SL with sharp interface, a SL with interface roughness, and corresponding alloy, respectively. It is worth noting that the predicted $\kappa$ decreases with increasing A and converges when A=10UC $\times$ 10UC. Herein, we choose A = 6 UC $\times$ 6 UC as the cross-sectional area for all simulations considering the limitation of the computational cost.

\section{Results and discussion}

\begin{figure}[htp]
	\centerline{\includegraphics[width=12cm]{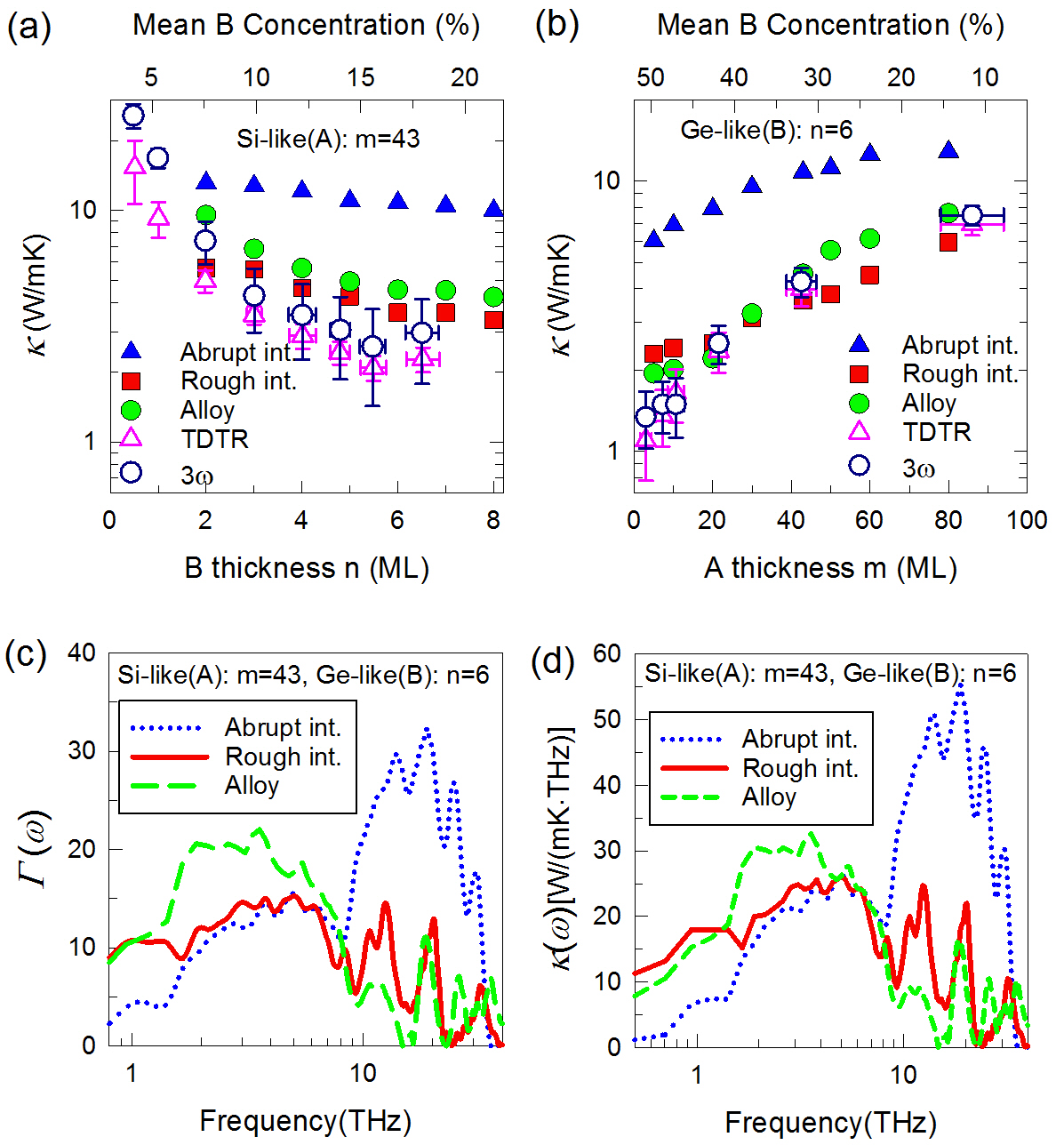}}
	\caption{\label{fig:3}
		The cross-plane thermal conductivity calculated at 30K for different $(A)_m/(B)_n$ structures corresponding to the case of the realistic $(Si)_m/(Ge)_n$ systems (a) as a function of B (Ge-like ``barrier'') layer thickness n with a fixed A (Si-like ``spacer'') layer thickness (m=43UC) and (b) as a function of A (Si-like ``spacer'') layer thickness m with a fixed B (Ge-like ``barrier'') layer thickness(n=6UC), respectively. Experimental results measured at 300K for $(Si)_m(Ge)_n$ SLs obtained by TDTR and 3$\omega$ methods from Ref.\cite{Chen2013} are also shown for comparison. In all cases, the period number N is 21. (c) Frequency dependent phonon transmission function and (d) spectral thermal conductivity of $(A)_{43}/(B)_6$ samples for the perfectly abrupt interface, rough interface ($\alpha =0.5$), and random alloy, respectively. }
\end{figure}

We start with calculating the cross-plane thermal conductivities of perfectly abrupt SLs and rough SLs driven by surface interdiffusion with variable B (corresponding to Ge-like ``barrier'') thickness (m ML), which are plotted in Fig.\ref{fig:3}(a). The thermal conductivity of a random alloy with the same B-concentration is also provided for comparison. It can be seen in all cases that $\kappa$ decreases with increasing thickness of the B layer, indicating that increasing thickness of the B layer is very effective in reducing $\kappa$. (This is reason why we name the material B as ``barrier''.) More importantly, our results show the cross-plane $\kappa$ of the SLs with interface roughness is always lower than the corresponding alloy values, which agrees well with experimentally measured results reported by Chen \cite{Chen2013}. Generally speaking, this intriguing result is ascribed to the surface-interdiffusion-driven intermixing around the A/B interfaces, which brings in multi-scale phonon scattering at all frequencies \cite{Chen2013,Chen2015}. Meanwhile, we also explore the effect of the thickness of the A (corresponding to Si-like ``spacer'') layers on the $\kappa$ as compared to experiment\cite{Chen2013}, as shown in Fig.\ref{fig:3}(b). Our calculation results show reasonable agreement with experimental data measured at 300K for realistic Si-Ge SLs \cite{Chen2013}, and lower $\kappa$ values for rough SLs are expected to achieve with increasing temperature. Interestingly, we note that the thermal conductivity of rough SLs can be reduced below the corresponding ``alloy limit'' only when the thickness of A layers increases up to 30UC ( for low B concentration ($\textless 30\%$) case), implying that the SL structures with different spacer thickness have different characteristic length of the phonon transport which are influenced by surface-interdiffusion-driven intermixing to different extend. Furthermore, the difference in the $\kappa$ between the rough SLs and random alloy remarkably increases with A layer thickness m. This suggests that the surface-interdiffusion-driven intermixing is more effective in SLs with large spacer thickness in achieving low $\kappa$ than with short one.

To understand the physical mechanism leading to the results above, we calculated the spectral phonon transmission functions for the perfectly abrupt interface, rough interface, along with the alloy, as depicted in Fig.\ref{fig:3}(c). It can be easily found that the transmission of low-to-mid frequency ($\textless 10$THz) phonons in the perfectly abrupt interface decreases significantly due to the interface scattering while strong alloy scattering in the corresponding alloy results in the much low transmission of high frequency ($\textgreater 10$THz) phonons. Compared with the perfect SLs, rough SLs weaken the effect of interface but bring alloy scattering, and thereby enhancing phonon scattering in all frequency ranges where phonons make a major contribution to the thermal conductivity. This is further confirmed by computing the spectral thermal conductivity of the corresponding structures plotted in Fig.\ref{fig:3}(d).

\begin{figure}[htp]
	\centerline{\includegraphics[width=12cm]{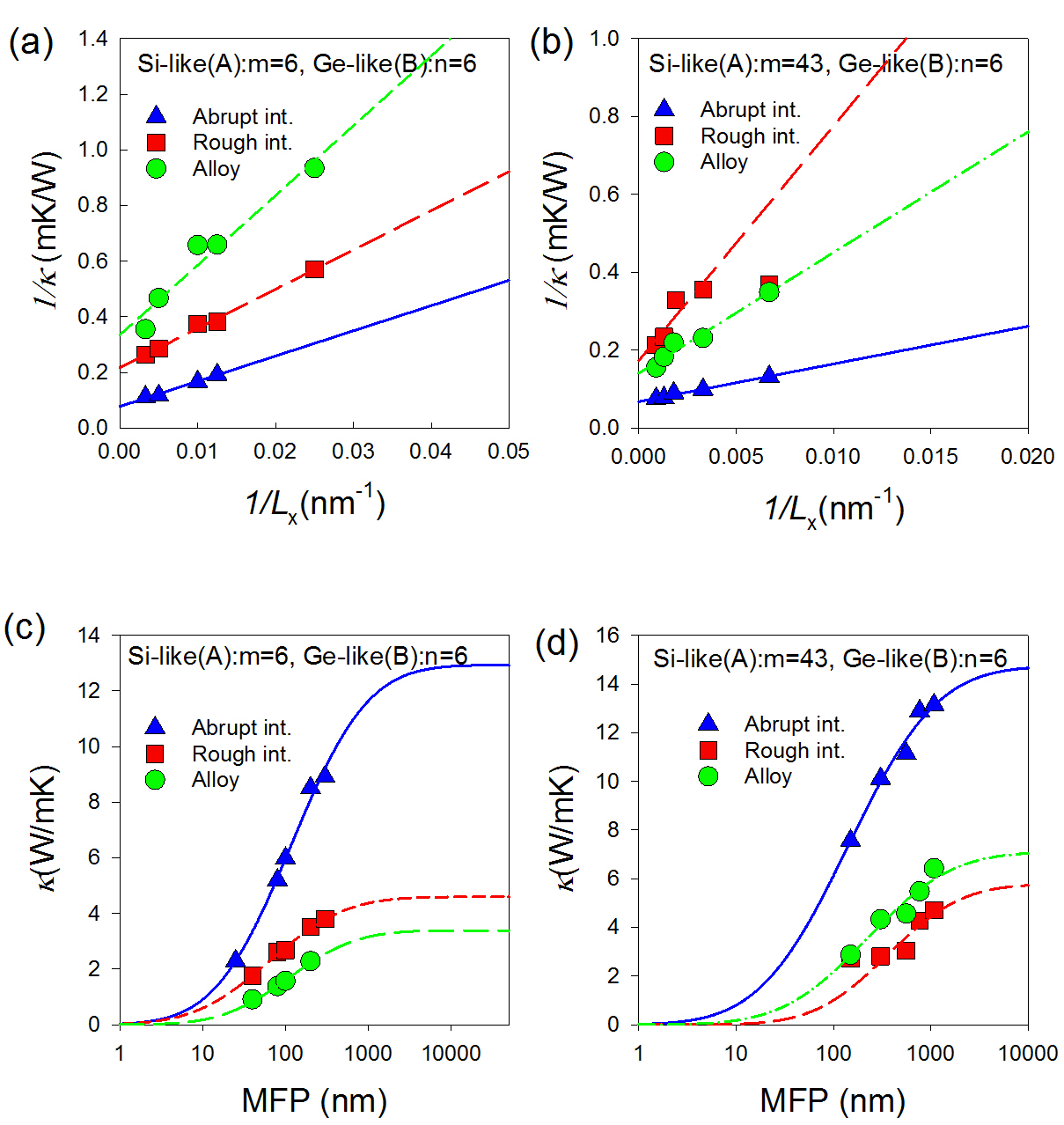}}
	\caption{\label{fig:4}
		Inverse of the predicted thermal conductivity as a function of inverse of the simulation cell length for $(A)_6/(B)_6$ samples (a) and for $(A)_{43}/(B)_6$ samples (b) corresponding to the realistic Si-Ge systems. In all cases, the thermal conductivity of infinitely long SLs ($\kappa_\infty$) in our simulations has been approximated by extrapolating to an infinite system size ($1/L_x \to 0$). Accumulation function of the thermal conductivity for $(A)_6/(B)_6$ samples (c) and for $(A)_{43}/(B)_6$ samples (d). }
\end{figure}

Next, we explore the dependence of cross-plane $\kappa$ on the sample length, in order to remove the constraints of a finite sample length in our NEMD simulations since coherent phonons with long mean free path are confined by the sample boundaries and contribute little to the overall $\kappa$ \cite{Schelling2002,Landry2008,Lampin2013}. The thermal conductivities of $(A)_{6}/(B)_6$ samples and $(A)_{43}/(B)_6$ samples corresponding to the realistic Si-Ge systems are shown in Fig.\ref{fig:4}(a-b) as a function of the sample length along thermal transport direction, for the perfectly abrupt interface, rough interface, and corresponding random alloy, respectively. According to $1/\kappa = 1/\kappa_\infty(1+\lambda/L_x)$ proposed by Schelling et al\cite{Schelling2002}, with $\lambda$ being the average of the mean free path (MFP) of dominant phonons, the thermal conductivity of infinitely long samples ($\kappa_\infty$) has been determined by extrapolating to an infinite system size ($1/L_x \to 0$). Using this linear extrapolation, we obtain the $\kappa_\infty$ at infinite cell length $L_x \to \infty$ to be 12.9329 W/mK (perfectly abrupt SLs), 3.3891 W/mK (rough SLs), and 2.9183 W/mK (alloy limit) for the sample with short period length ($(A)_6/(B)_6$), and to be 14.7298 W/mK (perfectly abrupt SLs), 5.7709 W/mK (rough SLs), 7.0962 W/mK (alloy limit) for the sample with long period length($(A)_{43}/(B)_6$). It can be seen that SLs with interface roughness is very effective in achieving the low $\kappa$ below the ``alloy limit'' for the sample with long period length, as mentioned in Fig.\ref{fig:3}(b). With the same method, we also obtain the MFP to be 117.4059nm (perfectly abrupt SLs), 90.0172nm (rough SLs), and 65.7337nm (alloy limit) for the sample with short period length ($(A)_6/(B)_6$), and to be 142.5578nm (perfectly abrupt SLs), 347.4317nm (rough SLs), and 220.0307nm (alloy limit) for the sample with short period length ($(A)_{43}/(B)_6$). To further investigate the spectral contribution of phonons to the thermal conductivity, Fig.\ref{fig:4}(c-d) shows the accumulation of thermal conductivity as a function of the phonon mean free path. We clearly see that the contribution to the thermal conductivity of the SL with short period length ($(A)_6/(B)_6$) largely originates from phonons of mean free path smaller than 100nm (corresponding to high frequency phonon) which are mainly dominated by alloy scattering. In contrast, the SL with long period length ($(A)_{43}/(B)_6$) has a much longer characteristic length of phonon transport than 100nm which are mainly influenced by interface scattering. This results further confirm the above points, and also explain why the thermal conductivity of rough SLs can break the ``alloy limit'' only when the A spacer thickness reaches up to a critical value.

\begin{figure}[htp]
	\centerline{\includegraphics[width=15cm]{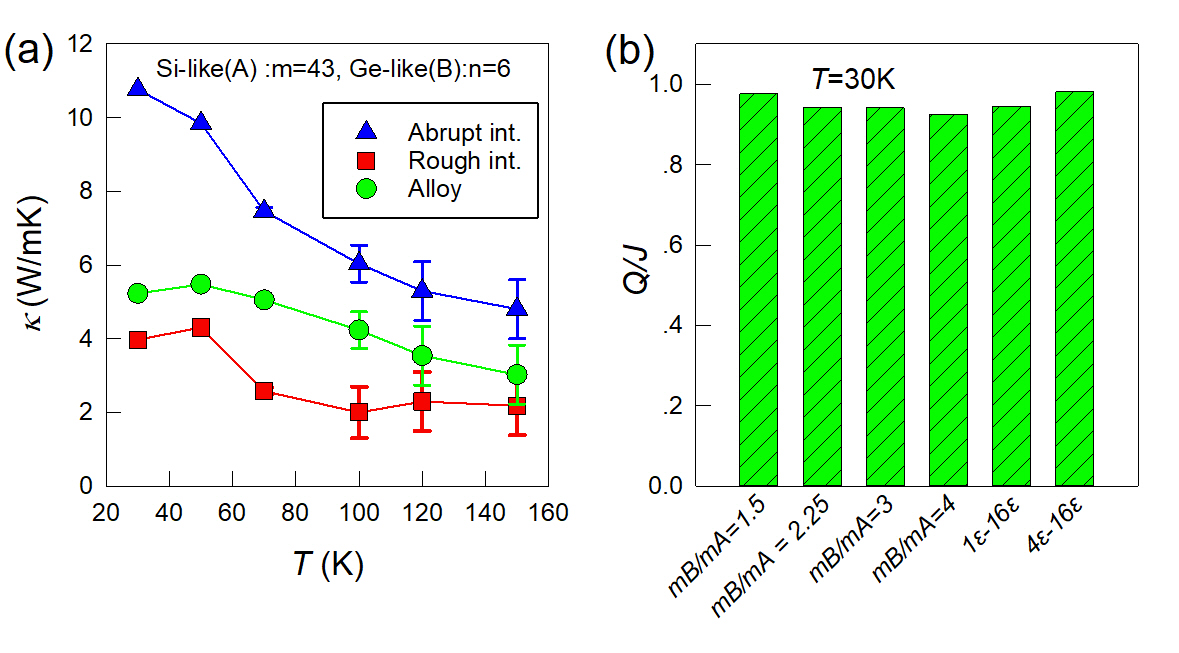}}
	\caption{\label{fig:5}
	(a) Thermal conductivities of $(A)_{43}/(B)_6$ samples corresponding to the realistic Si-Ge systems for the perfectly abrupt interface, rough interface ($\alpha =0.5$), and corresponding random alloy. (b) The ratio of the accumulation of spectral heat current $Q$ to the total heat flux $J$ for SLs materials with a broad span of atomic mass and bond strength.}
\end{figure}

Furthermore, we investigate the temperature dependence of $\kappa$. Within our potential set, we note that crystalline structures would be unsustainable beyond the maximum temperature of 150K. Fig.\ref{fig:5}(a) shows $\kappa$ as a function of temperature for a SL with sharp interface, a SL with interface roughness, and corresponding alloy. It can be found that $\kappa$ of SLs with rough interface is insensitive to temperature as the temperature rises, indicating that the phonon scattering for SLs is dominated by temperature-independent harmonic scattering caused by interface roughness, since the anharmonic scattering increases linearly with temperature and the specific heat is temperature independent based on classical MD. This further explains why $\kappa$ calculated at 30K can afford the experimental results measured at 300K. Therefore, surface-interdiffusion-driven intermixing can be expected to apply to broad temperature range for achieving the low $\kappa$ below the alloy limit.

\begin{figure}[htp]
	\centerline{\includegraphics[width=15cm]{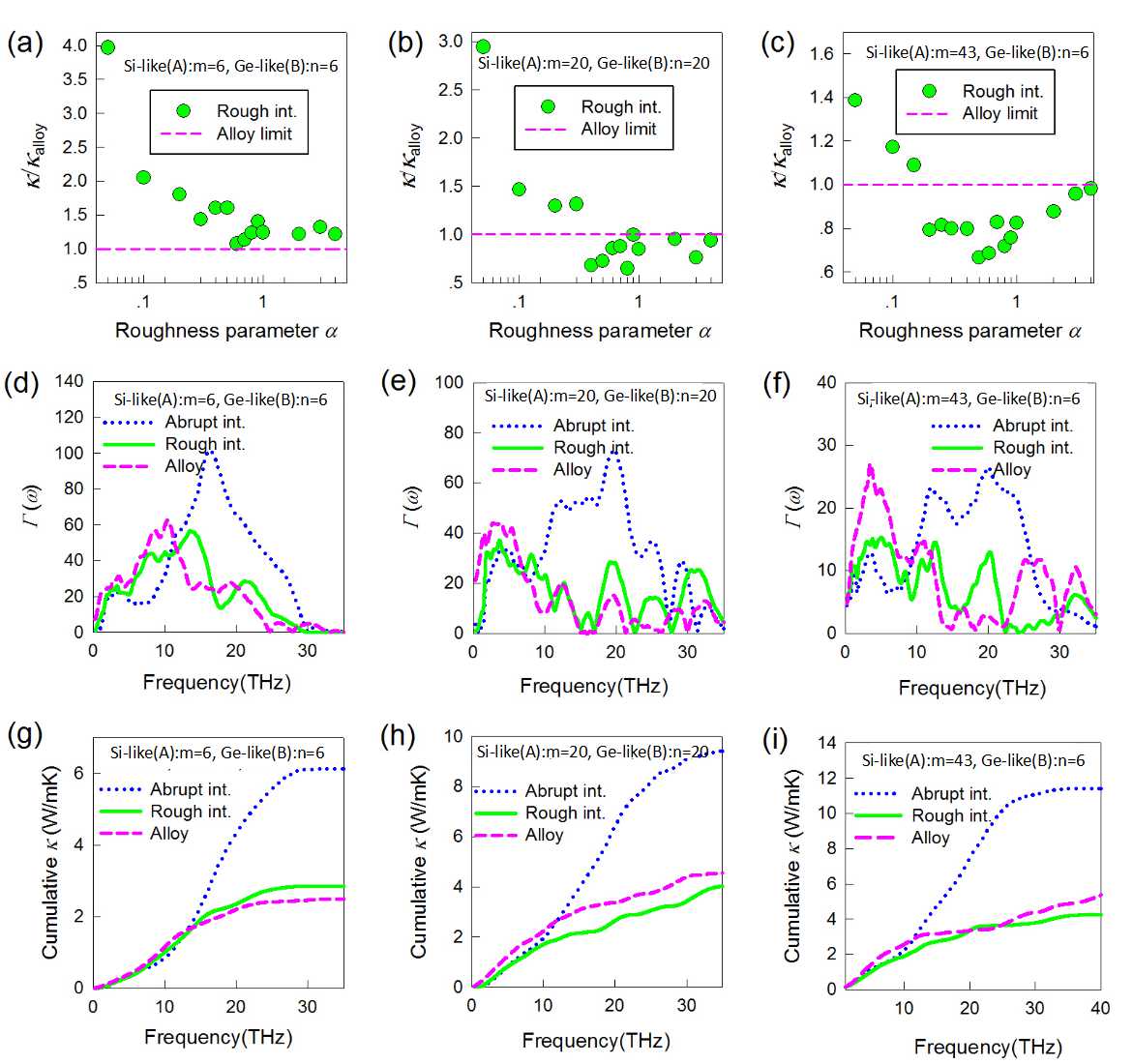}}
	\caption{\label{fig:6}
	The thermal conductivity relative to alloy limit $\kappa/\kappa_{alloy}$ as a function of interface roughness parameter $\alpha$ for (a) $(A)_6/(B)_6$ and (b) $(A)_{20}/(B)_{20}$ and (c) $(A)_{43}/(B)_6$ corresponding to the realistic Si-Ge systems. (d-f) Corresponding phonon transmission function changes with angular frequency for the perfectly abrupt interface, rough interface ($\alpha =0.5$), and random alloy. (g-i)Cumulative thermal conductivity as a function of mode angular frequency for different SL structures. }
\end{figure}

As discussed above, increasing intermixing driven by interdiffusion for a SL structure results in weakening of interface scattering and simultaneously enhancement of alloy scattering. This implies that the thermal conductivity of SLs must change with the interface roughness, and the interplay between the interface and alloy scattering should account for an optimal interface roughness which can minimize the thermal conductivity of SLs. To illustrate this effect, we calculated the relative thermal conductivity for rough SLs as compared to the alloy limit with increasing roughness, as shown in Fig.\ref{fig:6}(a-c). For a SL with uniform layer thickness $d_A=d_B=6$, we can see in Fig.\ref{fig:6}(a) that thermal conductivity $\kappa$ decreases monotonically with increasing interface roughness, and $\kappa$ gradually approaches to the corresponding alloy limit. In contrast to Fig.\ref{fig:6}(a), Fig.\ref{fig:6}(b) (for a SL with uniform layer thickness $d_A=d_B=20$UC) and Fig.\ref{fig:6}(c) (for a non-uniform SL with layer thickness $d_A=43$UC and $d_B=6$UC) show that $\kappa$ does not monotonically decrease with interface roughness. Instead, there exists an intermediate interface roughness which is capable to minimize the $\kappa$ value below the alloy limit (complete intermixing). To explain this difference, we calculated the phonon transmission function and cumulative thermal conductivity as a function of $\omega$ defined as the integrated $\kappa_{\mathrm{sp}}$ to $\omega$ for the corresponding SLs as interface roughness increases, as depicted in Fig.\ref{fig:6}(d-i). For all cases, we can see that as interface roughness increases, the phonon transmission function at low-to-mid frequencies ($\textless 10$THz) gradually increases while those of high frequencies ($\textgreater 10$THz) is suppressed more or less, suggesting that the surface-interdiffusion-driven intermixing leads to the weakening of interface scattering and the enhancement of alloy scattering, as discussed above. In contrast, for the SL with short period length as interface change from perfectly sharp to totally intermixed, the high frequency phonons ($\textgreater 10$THz) still keep a relatively large transmission function (see Fig.\ref{fig:6}(d)), which in fact make a dominant contribution to the thermal conductivity, while low-mid frequency phonons ($\textless 10$THz) contribute no more than $40\%$ of the $\kappa$ (see Fig.\ref{fig:6}(g)). This implies that the thermal conductivity of a SL with short period length is mainly dominated by the alloy scattering, which should be responsible for the results that thermal conductivity $\kappa$ decreases monotonically as interface roughness. Therefore, we believe that in this case alloy is more effective than the rough SL in achieving low thermal conductivity, as illustrated in Fig.\ref{fig:6}(a). In terms of the SL with large period thickness, whether uniform or not, we find that alloy scattering due to surface-interdiffusion-driven can lead to a significant reduction of transmission function of high frequency phonons ($\textgreater 10$THz), meanwhile, phonon transmission function at low-to-mid frequency ($\textless 10$THz) ranges increases at a level already close to alloy limit (see Fig.\ref{fig:6}(e-f)). This implies that the thermal conductivity of the SLs with large period length is mainly influenced by interface scattering as interface roughness increases. This is further illustrated by Fig.\ref{fig:6}(h-i), which display the accumulative thermal conductivity of the SLs with sharp interface, rough SLs, along with random alloy, respectively. We can see the contribution to the $\kappa$ from low-to-mid frequency phonons exceeds $50\%$ for both cases. This intriguing results demonstrate the significant effect of surface-interdiffusion-driven intermixing on thermal transport for the SL sample with large period length, and it opens the door to achieve the low $\kappa$ much below the corresponding ``alloy limit'' for realizing high performance thermoelectrics in SL structures.

\begin{figure}[htp]
	\centerline{\includegraphics[width=15cm]{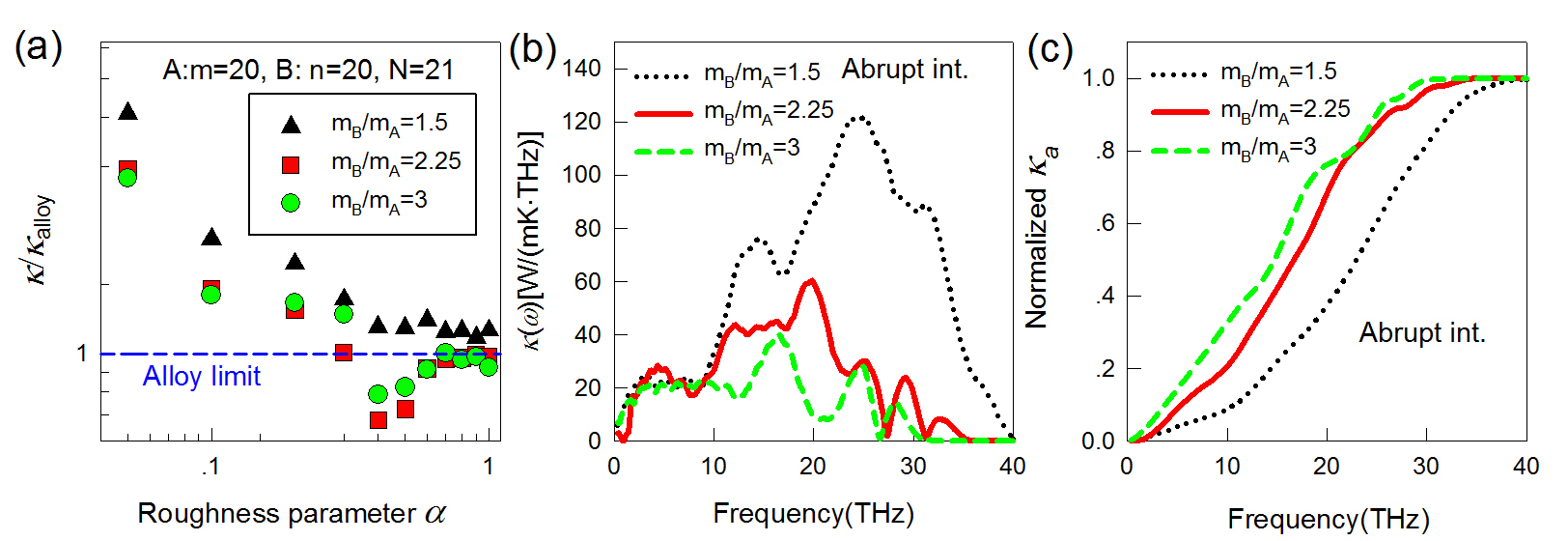}}
	\caption{\label{fig:7}
	(a) The cross-plane thermal conductivity of $(A)_{20}/(B)_{20}$ SL relative to the alloy limit ( $\kappa/\kappa_{alloy}$) as a function of interface roughness parameter $\alpha$ for different mass mismatch. (b) Spectral thermal conductivity for the SL with perfect interface. (c) Cumulative thermal conductivity normalized by the total $\kappa$ as a function of mode angular frequency for different SL structures. }
\end{figure}

From applied point of view, it is also worth exploring the effect of interface roughness driven by interdiffusion on thermal conductivities for SLs materials with a broad span of atomic mass and bond strength, i.e, AlAs/GaAs, Bi$_2$Te$_3$/Sb$_2$Te, which have very different phonon properties. We first study three systems with different mass ratio. The first system is composed of two materials with atomic mass of $m_A=40$g/mol and $m_B=60$g/mol, respectively, and hence denoted as $m_B/m_A=1.5$. Accordingly, the second system with $m_A=40$g/mol and $m_B=90$g/mol masses is denoted as $m_B/m_A=2.25$, and the third one with $m_A=40$g/mol and $m_B=120$g/mol masses is denoted as $m_B/m_A=3$. Fig.\ref{fig:7}(a) depicts the value of $\kappa$ relative to the corresponding alloy limit for the three systems as interface change from perfectly abrupt to totally intermixed. It can be seen that the surface-interdiffusion-driven intermixing is more effective for materials with larger mass mismatch ratio ($m_B/m_A=2.25$ and $m_B/m_A=3$) with a view to achieving much lower $\kappa$ than alloy limit. The reason is that as the mass ratio increases the contribution to the $\kappa$ from high frequency phonons ($\textgreater 10$THz) significantly decreases due to the effect of phonon softening (see Fig.\ref{fig:7}(b)), and therefore, interface scattering to low-mid frequency phonons gradually plays a dominant role. It can be further verified from Fig.\ref{fig:7}(c), the contributions of dominant phonons to the $\kappa$ gradually shift towards into lower frequency ranges with increasing mass mismatch.

\begin{figure}[htp]
	\centerline{\includegraphics[width=9cm]{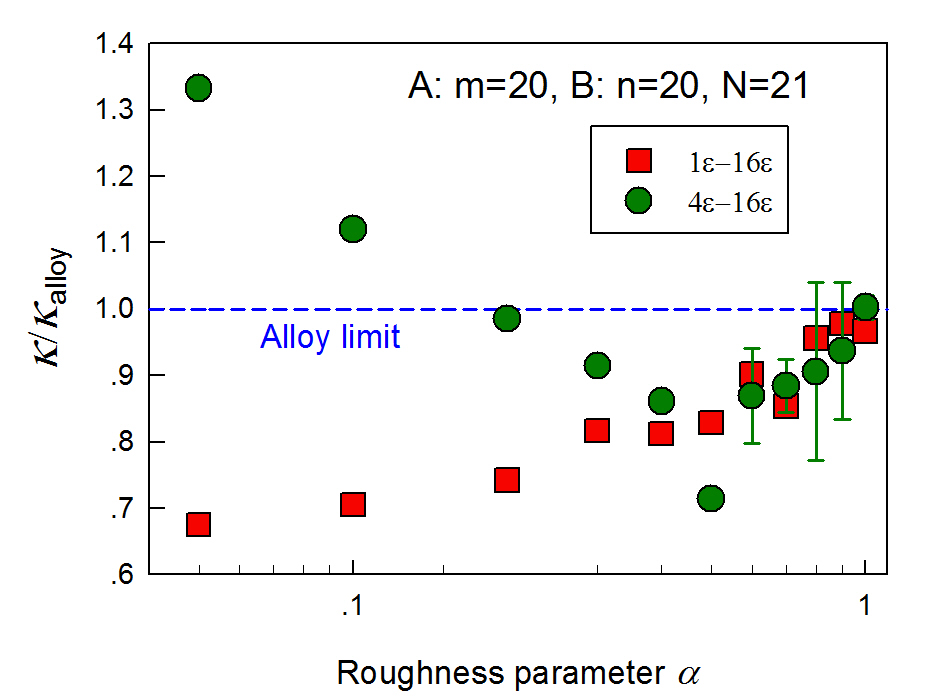}}
	\caption{\label{fig:8}
	(a) The cross-plane thermal conductivity of $(A)_{20}/(B)_{20}$ SLs relative to the alloy limit ($\kappa/\kappa_{alloy}$) as a function of interface roughness parameter $\alpha$ for different bond strength mismatch. Error bars were calculated based on three independent simulations.}
\end{figure}

\begin{figure}[htp]
	\centerline{\includegraphics[width=13cm]{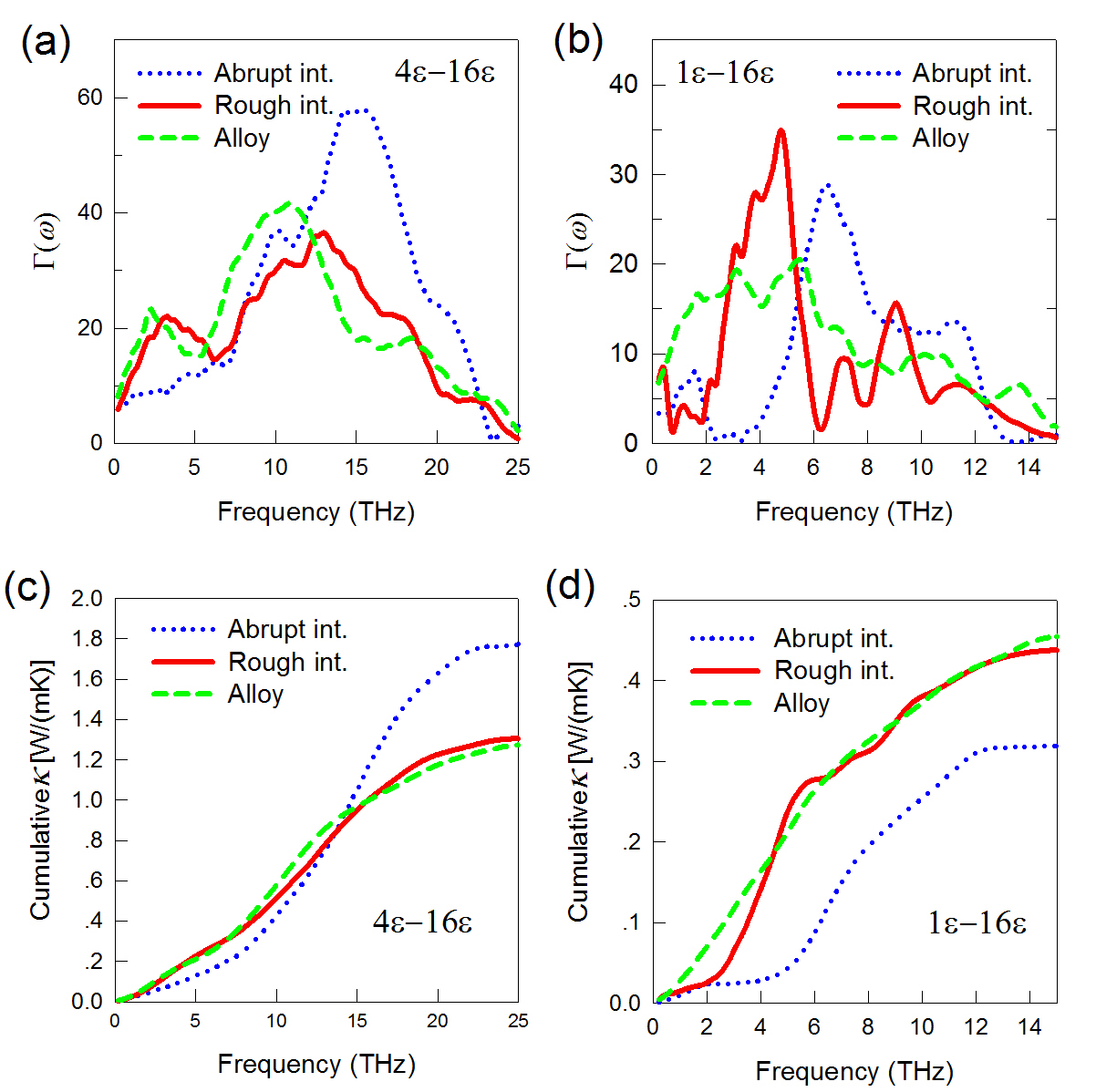}}
	\caption{\label{fig:9}
	(a-b) The spectral thermal conductivity (c-d) Cumulative thermal conductivity of $(A)_{20}/(B)_{20}$ SL with sharp interface, SL with interface roughness, and random alloy for different bond strength mismatch. }
\end{figure}

We also study two system with different bond strength ratio. The first system consists of two base materials of which the $\epsilon$ is 1 and 16 times of the $\epsilon$  for argon, and, accordingly, denoted as $1\epsilon$ - $16\epsilon$. Similarly, the second system of which the $\epsilon$ is 4 and 16 times of that for argon is denoted as $4\epsilon$ - $16\epsilon$. In both systems, all the atoms have a mass of 40 g/mol. In Fig.\ref{fig:8}, it is obvious that the evolutions of the $\kappa$ for rough SLs with mismatch ratio $1\epsilon$ -$16\epsilon$ and with mismatch ratio $4\epsilon$ -$16\epsilon$ present two completely reverse trend as interfaces change from perfectly abrupt to totally intermixed. In SLs with small bond strength mismatch ratio ($4\epsilon$ -$16\epsilon$), there exist an optimal interface roughness which can minimize the $\kappa$ below the corresponding alloy, while the $\kappa$ increases monotonically with interface roughness in SLs with much mismatch ratio($1\epsilon$ -$16\epsilon$). Comparing Fig.\ref{fig:9}(a) and Fig.\ref{fig:9}(b), which display the transmission function for the system with small bond strength mismatch ratio and the system with much bond strength mismatch ratio, respectively, we observe that for the latter one, surface-interdiffusion-driven intermixing has a much more significant effect on the transmission function in low-to-mid frequency phonons ($\textless 6$ THz). It means that, for the latter one, the $\kappa$ of rough SLs is mainly dominated by the weakening of the interface scattering stemming from interdiffusion. Taking cumulative thermal conductivity of a SL with perfectly abrupt interface as a reference, we see in Fig.\ref{fig:9}(c-d) that for the SL with much bond strength mismatch surface-interdiffusion-driven intermixing significantly weakens the effect of phonon scattering in the low-to-mid frequency ranges ($\textless 6$THz), which accounts for the result that the $\kappa$ monotonically increases with interface roughness. This indicates that a large lattice mismatch in the ideally SLs results in materials with $\kappa$ much below the ``alloy limit''. A similar trend was also reported for SLs with a large mass difference in the two intercalated layers, or weakened interactions between layers\cite{Mizuno2015,Guo2015Approaching}.

\section{Conclusions}

To summarize, inspired by experiments, we performed NEMD simulations to investigate the effect of surface-interdiffusion-driven intermixing on the thermal transport through SLs. We calculated the cross-plane thermal conductivity of SLs as the interface changes from perfectly abrupt to totally intermixed. Our simulation results suggest that surface-interdiffusion-driven intermixing is crucial for lowering the thermal conductivity $\kappa$, especially, when the period thickness of A (``spacer'') layers increases up to a critical value the $\kappa$ of rough SLs is capable to break the corresponding ``alloy limit''. This is attributed to the interplay between the alloy and interface scattering. More importantly, we find that $\kappa$ of SLs does not monotonically changes with interface roughness for the SLs with large period length. Instead, there exists an intermediate interface roughness which can minimize the thermal conductivity. In addition, we also demonstrate the significant effect of interdiffusion on thermal transport for SLs materials with a broad span of atomic mass and bond strength. Surface-interdiffusion-driven intermixing is found to be more effective in achieving the low $\kappa$ below the alloy limit for SL materials with large mass mismatch than with small one. More interestingly, we find it possible for SLs materials with large lattice mismatch (i.e., bond strength) to design an ideally sharp interface structure with $\kappa$ much below the ``alloy limit''. These findings provide guidance for realizing high efficiency thermoelectrics by optimizing the thermal conductivity in SL structures.

\section*{Acknowledgements}
We thank Professor J. Li at MIT for helpful discussions. W.L. acknowledges support from Natural Science Foundation of Guangdong Province under Grant No.2017A030310377. Simulations are performed on the ${Huashan}^{\#}$1 HPC cluster in State Key Laboratory for Mechanical Behavior of Materials and ${Huashan}^{\#}$2 HPC cluster in Center of Microstructure Science (CMS), Xi'an Jiaotong University, China.

%\bibliography{refs}

\end{document}